\documentclass[12pt]{article}

 \usepackage[makeroom]{cancel}
 \usepackage{epstopdf}
 \usepackage{amssymb, latexsym, amsmath}
 \usepackage[dvips]{color}
 \usepackage{graphicx}
 \usepackage{color}
 \usepackage{deluxetable}
  \usepackage{placeins}
\usepackage{float}
\usepackage{lscape}

  \usepackage{ifpdf}
  \usepackage{epsfig}
  \usepackage[latin1]{inputenc}
  \usepackage{amsfonts}

  \usepackage{amsbsy}
  \usepackage{amstext}
  \usepackage{amsopn}
  \usepackage{amsgen}

 \newcommand{\inc}{{\it i}}

 \newcommand{\be}{\begin{equation}}
 \newcommand{\ee}{\end{equation}}
 \newcommand{\ba}{\begin{eqnarray}}
 \newcommand{\ea}{\end{eqnarray}}
 \newcommand{\bs}{\begin{subequations}}
 \newcommand{\es}{\end{subequations}}

 \newcommand{\erbold}{\mbox{{\boldmath $
 r$}}}

  \newcommand{\eRbold}{\mbox{{\boldmath $
  {R}$}}}
  \newcommand{\Rbold}{\mbox{{\boldmath $
  {R}$}}}

\begin{document}

 %
 %

  \title{
           ${{~~~~~~~~~}^{^{^{
         ~~~~~~~~~~~~~~~~~The~
                 Astronomical~Journal
        \,~150~:~98\,~(2015)
                  }}}}$
                  ~
 {\Large{\textbf{Tidal Evolution of Asteroidal Binaries.\vspace{2mm}\\
                 Ruled by Viscosity. ~Ignorant of Rigidity.}
            }}}
 \author{
 ~\\
 {\Large{Michael Efroimsky}}\\
 {\small{US Naval Observatory, Washington DC 20392 USA}}\\
 {\small{e-mail: michael.efroimsky$\,$@$\,$usno.navy.mil~}}\\
 ~\\
  }

 %
 %
 \maketitle
 %
 %

     \date{}

 \begin{abstract}
 ~\\
 This is a pilot paper serving as a launching pad for study of orbital and spin evolution of binary asteroids.
 The rate of tidal evolution of asteroidal binaries is defined by the dynamical Love numbers $\,k_l\,$ divided by quality factors $\,Q\,$. Common in the literature is the (oftentimes illegitimate) approximation of the dynamical Love numbers with their static counterparts. Since the static Love numbers are, approximately, proportional to the inverse rigidity, this renders a popular fallacy that the tidal evolution rate is determined by the product of the rigidity by the quality factor: $\,k_l/Q\propto 1/(\mu Q)\,$. In reality, the dynamical Love numbers depend on the tidal frequency and all rheological parameters of the tidally perturbed body (not just rigidity). We demonstrate that in asteroidal binaries the rigidity of their components plays virtually no role in tidal friction and tidal lagging, and thereby has almost no influence on the intensity of tidal interactions (tidal torques, tidal dissipation, tidally induced changes of the orbit). A key quantity that overwhelmingly determines the tidal evolution is a product of the effective viscosity $\,\eta\,$ by the tidal frequency $\,\chi\,$. The functional form of the torque's dependence on this product depends on who wins in the competition between viscosity and self-gravitation. Hence a quantitative criterion, to distinguish between two regimes. For higher values of $\,\eta\chi\,$, we get $\,k_l/Q\propto 1/(\eta\chi)\;$; $\,$while for lower values we obtain $\,k_l/Q\propto \eta\chi\,$. Our study rests on an assumption that asteroids can be treated as Maxwell bodies. Applicable to rigid rocks at low frequencies, this approximation is used here also for rubble piles, due to the lack of a better model. In the future, as we learn more about mechanics of granular mixtures in a weak gravity field, we may have to amend the tidal theory with other rheological parameters, ones that do not show up in the description of viscoelastic bodies. This line of study provides a tool to exploring the orbital history of asteroidal pairs, as well as of their final spin states.
 ~\\
 \end{abstract}

 \pagebreak

 \section{Dynamics of binary asteroids}

 Numerous binary asteroids have been discovered by now (see, e.g., Pravec \& Harris 2007, Taylor \& Margot 2011, Jacobson \& Scheeres 2011, Margot et al. 2015). Analysis of their orbital and rotational dynamics will give us a key to understanding these asteroid's internal properties from observation of their orbits and spin.

 Spin-orbit evolution of a binary is implemented through tidal torques that the companions exert upon one another. Torques due to the permanent figure also participate in this interaction and play a role in tidal capture into spin-orbit resonances (e.g., Noyelles et al. 2014, Batygin \& Morbidelli 2015). In this paper, however, we shall concentrate on tides.

 \subsection{Difficulties in modeling of binary asteroids' tidal dynamics}

 At first glance, the strategy and methods for exploring binary asteroids' dynamics should be the same as those employed in the studies of planets and moons. In reality, we are facing a more complicated case, for three main reasons:
 \begin{itemize}
 \item[1.~] Many asteroidal pairs are very close, so calculation of their tidal interaction requires inclusion of terms higher than quadrupole
     (Mathis \& Le Poncin-Lafitte 2009, Comp{\`{e}}re \& Lema{\^{i}}tre 2014,
     Taylor \& Margot 2010).
 \item[2.~] Tidal reaction of a connected (not rubble) small body is overwhelmingly defined by rheology and only to a minuscule degree by self-gravitation (which is not the case for planets -- see, e.g., Efroimsky 2012$\,$b).
 \item[3.~] Many asteroids, however, are believed to be loose aggregates. Their rheology is largely defined by microgravity, and may differ from the rheology of solids not only in parameters' values. A constitutive equation for granular matter in a weak gravity field may, in principle, contain parameters other than rigidity or viscosity.
 \end{itemize}

 \subsection{Tidal torques}

 When an extended celestial body is deformed by the potential of an exterior perturber, the distortion results in an additional tidal potential produced by the extended body itself. The latter potential acts back on the perturber and changes its orbital motion.

 In each point of the extended body, the exterior perturbing potential changes in time and usually contains an infinite set of Fourier harmonics. We shall assume the reaction linear, so the Fourier spectrum of the reaction (i.e., of the shape changes and of the resulting tidal potential) mimics the spectrum of the perturbing potential.

 As ever in physics, the reaction lags behind action. Consequently, one may expect that the Fourier modes of the tidal elevation and tidal potential should always lag behind the corresponding modes of the perturbing potential. This point, however, needs attention: falling behind in time does $\,${\it{not}}$\,$ fix the sign of a geometric lag. \footnote{~Causality requires reaction to fall behind action, $\,${\it{in time}}$\,$. So each Fourier mode of the additional tidal potential falls behind the appropriate mode of the perturbing potential, by a time lag $\,\Delta t\,$ which is mode-dependent but is always positive definite. This, however, does not imply that the $\,${\it{phase}}$\,$ lags should always be positive. Indeed, the phase lag at a certain Fourier mode is a product of this mode by the appropriate time lag (see the equation \ref{lags} below), so the sign of this lag always coincides with that of the mode. Likewise, the observable geometric lags (related to the phase lag through the formula \ref{geometric} below), share the sign with the Fourier mode. Accordingly, no matter whether the perturber is below or above the synchronous orbit, some components of the tide, geometrically, fall behind the perturber; and some lead. For example, the principal (semidiurnal) Phobos-raised bulge in Mars falls behind. Of the lesser bulges, some fall behind and some lead, at various angular rates.}

 Since the perturber ``feels" the tides it itself creates in the extended body, the perturber experiences an orbital tidal force and an orbital tidal torque. Then the deformed body experiences a tidal torque of an opposite sign, which acts to change this body's spin. This interaction is reciprocal, in that the deformed body also acts as a perturber upon its companion -- and exerts a despinning torque thereon, and experiences an opposite orbital torque.

 \subsection{A specific problem}

 In the literature on tidal dynamics, it has long become a convention to introduce the Love numbers and quality factors as separate entities. More often than not, authors
 postulate some frequency-dependence of the quality factors, treating the Love numbers as constant parameters (e.g., Mignard 1979, 1980, 1981; Correia \& Laskar 2004; Efroimsky \& Lainey 2007; Cheng et al. 2014). Above that, the dynamical Love numbers are very often identified with their static counterparts (Goldreich \& Peale 1968, Peale \& Cassen 1978, Murray \& Dermott 1999, Taylor \& Margot 2010).

 The static Love numbers are, approximately, proportional to the inverse rigidity.  For example, the static quadrupole Love number is given by
   \ba
   k^{\textstyle{^{(static)}}}_{2}\,=\;\frac{3}{2}\;\,\frac{1}{1\;+\;\frac{\textstyle 19~\mu}{\textstyle 2~\mbox{g}\,\rho\,R}}~\approx~\frac{3}{19}\;\frac{\,\textstyle \mbox{g}~\rho~R\,}{\textstyle \mu}~~~,
   \label{}
   \ea
 with g, $\,\rho\,$, and $\,R\,$ being the surface gravity, density and radius of a spherical body. This altogether renders a common misunderstanding that the tidal evolution rate is determined by the product of the rigidity by the quality factor: $\,k_l/Q\propto 1/(\mu Q)\,$.

 A minor issue with this treatment is that the tidal quality factors, as functions of frequency, are different for different values of the degree $\,l\,$ (and are different from the seismic quality factor). Negligible at ordinary frequencies, these differences become critical in the zero-frequency limit, i.e., at crossing of spin-orbit resonances (Makarov et al. 2012, Noyelles et al. 2014).

 A bigger issue -- the one to be dealt with -- is that the dynamical Love numbers, aside from their frequency dependence, depend on all rheological parameters of the tidally perturbed body (not just rigidity). \footnote{~The Andrade model serves as an example of rheology containing parameters other than rigidity and viscosity -- see the equations (\ref{LL44}) in Section \ref{section3.2} below. More examples of the kind can be found, e.g., in Henning et al. (2009).}
 Our goal is to demonstrate that in asteroidal binaries the rigidity of their components plays virtually no role in tidal friction and tidal lagging. Consequently, it has virtually no influence on tidal despinning, or tidal heating, or tidally induced orbital evolution. As we shall see, a parameter that overwhelmingly determines the tidal torques (and therefore dissipation, as well as orbit changes) is a product of the effective viscosity $\,\eta\,$ by the tidal frequency $\,\chi\,$.

 \section{Notation and formalism}

 \FloatBarrier
 \begin{deluxetable}{lr}
 \tablecaption{Symbol key \label{Table}}
 \tablewidth{0pt}
 \tablehead{
 \multicolumn{1}{c}{Notation}  &
 \multicolumn{1}{c}{Description}
 }
   \startdata
 $\theta$ & \dotfill the rotation angle of the body\\
 $R$ & \dotfill the radius of the body \\
 ${\cal{T}}^{\rm{^{\,(TIDE)}}}_z$ & \dotfill the polar component of the tidal torque acting on the body\\
 $r$ & \dotfill the instantaneous distance between the body and the perturber \\
 $a$ & \dotfill the semimajor axis \\
 $e$ & \dotfill the orbital eccentricity \\
  $i$ & \dotfill the orbital inclination (obliquity) \\
   $\Omega$ & \dotfill the longitude of the node \\
      $\omega$ & \dotfill the argument of the pericentre \\
 ${\cal{M}}$ & \dotfill the mean anomaly  \\
  $n$ & \dotfill the mean motion ~\\
 $G$ & \dotfill Newton's gravitational constant \\
   $\omega_{\textstyle{_{lmpq}}}$ & \dotfill a tidal Fourier mode \\
   $\chi_{\textstyle{_{lmpq}}}$ & \dotfill a forcing tidal frequency \\
   $\chi$ & \dotfill a shorter notation for $\,\chi_{\textstyle{_{lmpq}}}\,$ \\
 $\tau_{_M}$ & \dotfill the viscoelastic characteristic time (Maxwell time) \\
 $\tau_{_A}$ & \dotfill the inelastic characteristic time (Andrade time)\\
 $\eta$ & \dotfill the viscosity \\
 $\mu$ & \dotfill the unrelaxed rigidity modulus \\
 $J$ & \dotfill the unrelaxed compliance \\
 $\alpha$ & \dotfill the Andrade parameter \\
 \enddata
  \end{deluxetable}

 Referring the reader to Appendix \ref{AppendixA} for more detail, here we provide basic formulae to be used.
 In Table 1 and thereafter, the word $\,${\it{body}}$\,$ will mean: $\,${\it{the tidally perturbed body}}$\,$. Its companion will be termed as the $\,${\it{perturber}}$\,$. Evidently, their roles are always interchangeable, and the overall dynamics is defined by tides in both companions.

 \subsection{Modes and frequencies\label{section2.1}}

 Both the perturbing potential and the response (tidal distortion of the perturbed body and the resulting additional tidal potential) can be expanded over the Fourier modes
 \ba
 \omega_{\textstyle{_{lmpq}}}~=~(l-2p)\;\dot{\omega}\,+\,(l-2p+q)\;n\,+\,m\;(\dot{\Omega}\,-\,\dot{\theta})
 ~\approx~(l-2p+q)\;n\,-\,m\;\dot{\theta}
 \label{omega}
 \label{A1}
 \ea
 parameterised with four integers $\,l$, $m$, $p$, $q\,$ (Kaula 1964). Here $\,\theta\,$ and $\,{\bf{\dot{\theta\,}}}\,$ are the rotation angle and rotation rate of the tidally perturbed body, while $\,\omega\,$ and $\,\Omega\,$ are the perturber's argument of the pericentre and the longitude of the node, as seen from the perturbed body. The ``anomalistic" mean motion $\,n\,$ is $\,${\it{defined}}$\;$ through $\,n\equiv{\bf{\dot{\cal{M}}}}\,$ (with $\,{\cal{M}}=\,0\,$ in the pericentre). The modes $\,\omega_{\textstyle{_{lmpq}}}\,$ can assume either sign, while the physical forcing frequencies in the distorted body are positive definite:
 \ba
 \chi_{\textstyle{_{lmpq}}}\,=\,|\,\omega_{\textstyle{_{lmpq}}}\,|~\approx~|\,(l-2p+q)\;n\,-\,m\;\dot{\theta}\,|\,~,
 \label{chi}
 \label{A2}
 \ea
 so
 \ba
 \omega_{\textstyle{_{lmpq}}}~=~\chi_{\textstyle{_{lmpq}}}~\,\mbox{Sgn}\,(\omega_{\textstyle{_{lmpq}}})~~.
 \label{A3}
 \ea

 \subsection{The despinning tidal torque}

 Development of the polar torque can be found, e.g., in Efroimsky (2012$\,$a). The torque contains both an oscillating and a secular part, the latter  looking as
 \ba
 \nonumber
 \langle\,{\cal{T}}_z^{\rm{^{\,(TIDE)}}}\rangle\;= \qquad\qquad\qquad\qquad\qquad\qquad\qquad\qquad\qquad\qquad\qquad\qquad\qquad\qquad\qquad\qquad
 ~\\
 \label{4}\\
 \nonumber
 \,2\,G\,M^{{{\,2}}}
 \sum_{{\it{l}}=2}^{\infty}
 \frac{R^{\textstyle{^{2l\,+\,1}}}}{
 a^{\textstyle{^{2l\,+\,2}}}}
 \sum_{m=0}^{l}
 \frac{(l-m)!}{(l+m)!}\;m
 %
 %
  \sum_{p=0}^{l}F^{\,2}_{lmp}(i)
 \sum^{\it \infty}_{q=-\infty}
 G^{\,2}_{lpq}(e)\;k_l(\omega_{\textstyle{_{lmpq}}})\;\sin\epsilon_l(\omega_{\textstyle{_{lmpq}}})\,~.~
 \ea
 Here $\,G\,$ is Newton's gravity constant; $\,M\,$ is the mass of the perturber; $\,a,\,i,\,e\,$ are the semimajor axis, inclination, and eccentricity; and the angular brackets $\,\langle\,.\,.\,.\,\rangle\,$ denote time averaging. The notations $\,F_{lmp}(i)\,$ and $\,G_{lpq}(e)\,$ signify the inclination functions and the eccentricity polynomials. The degree-$l\,$ Love numbers $\,k_l(\omega_{lmpq})\,$ and phase lags $\,\epsilon_{\textstyle{_l}}(\omega_{lmpq})\,$ are functions of the Fourier tidal modes (\ref{omega}) -- see Appendix \ref{AppendixA}.

 In the series (\ref{4}), some terms are negative (despinning) and some are positive (spinning-up). Usually, the leading term is despinning and so is the overall torque. This, however, is not obligatory. For example, when the initial spin is prograde but slower than the mean motion, or when the initial spin is retrograde, the tidal torque acts to increase the spin rate in the prograde sense (Noyelles et al. 2014). Also mind that capture into spin-orbit resonances takes place due to a combined action of the tidal torque and a torque caused by the permanent triaxiality. Entrapment into higher than 1:1 spin-orbit resonances is more probable for tighter binaries and higher eccentricities.

  \subsection{The orbital tidal torque}

 The orbital tidal torque acting on the perturber is opposite to the torque (\ref{4}) wherewith the perturber despins or spins up the tidally distorted body. This indicates that the products $\,k_l/Q_l\,=\,k_l(\omega_{\textstyle{_{lmpq}}})\,\sin\epsilon_l(\omega_{\textstyle{_{lmpq}}})\,$ emerging in (\ref{4}) must be key players not only in the spin evolution but also in orbital evolution.

 In practice, orbital calculations employ not the orbital torque, but the tidal potential. As distinct from the torque, the potential carries the factors  $\,k_l(\omega_{\textstyle{_{lmpq}}})\,\cos\epsilon_l(\omega_{\textstyle{_{lmpq}}})\,$. Recall, however, that the potential is inserted into the Lagrange- or Delaunay-type planetary equations. There it is differentiated with respect to orbital elements, and we once again end up with \footnote{~In the planetary equations, we also get small terms with $\,k_l(\omega_{\textstyle{_{lmpq}}})\,\cos\epsilon_l(\omega_{\textstyle{_{lmpq}}})\,$.
 They emerge because the tidally deformed body is no longer a perfect sphere, and its mass cannot be assumed to be concentrated at its centre. As a result of this, the anomalistic
 mean motion $\,n\equiv\,\stackrel{\bf{\centerdot}}{\cal{M\,}}$ slightly exceeds the Keplerian mean motion $\,\sqrt{G\,(M+m)/a^3\,}\,$.
 The terms with $\,k_l(\omega_{\textstyle{_{lmpq}}})\,\cos\epsilon_l(\omega_{\textstyle{_{lmpq}}})\,$ do not show up in the planetary equations for $\,a\,$, $\,e\,$, $\,i\,$, ~and cause only small secular changes in $\,{\cal{M}}\,$, $\,\omega\,$ and $\,\Omega\,$.
 }
 $\;k_l(\omega_{\textstyle{_{lmpq}}})\;\sin\epsilon_l(\omega_{\textstyle{_{lmpq}}})\,$.

 Although our paper addresses $\,k_l/Q_l\,=\,k_l(\omega_{\textstyle{_{lmpq}}})\,\sin\epsilon_l(\omega_{\textstyle{_{lmpq}}})\,$,
 for completeness we also provide in Appendix \ref{App} an expression for $\,k_l(\omega_{\textstyle{_{lmpq}}})\,\cos\epsilon_l(\omega_{\textstyle{_{lmpq}}})\,$, for the Maxwell body and for some other rheologies.

 \section{Formalism to describe rheology}

 As explained in Appendix \ref{AppendixA}, in an $\,lmpq\,$ term of the torque we can switch from the Fourier mode $\,\omega_{\textstyle{_{lmpq}}}\,$ to the physical frequency $\,\chi_{\textstyle{_{lmpq}}}\,=\,|\,\omega_{\textstyle{_{lmpq}}}\,|\;\,$:
 \ba
 k_{\textstyle{_l}}(\omega_{\textstyle_{lmpq}})~\sin\epsilon_{\textstyle{_l}}(\omega_{\textstyle_{lmpq}})
 \,=\, k_{\textstyle{_l}}(\chi_{\textstyle_{lmpq}})~\sin\epsilon_{\textstyle{_l}}(\chi_{\textstyle_{lmpq}})\;\mbox{Sgn}\,\omega_{\textstyle_{lmpq}}
 \quad.
 \label{}
 \ea
 The functional form of the dependence of $\,k_{\textstyle{_l}}(\chi)~\sin\epsilon_{\textstyle{_l}}(\chi)\,$ upon the frequency $\,\chi=\chi_{\textstyle{_{lmpq}}}\,$ is defined by self-gravitation of the body and by its rheological properties.

  It is common in the literature to denote $\,\sin\epsilon_{\textstyle{_l}}\,$ with $\,1/Q_{\textstyle{_l}}\,$, $\,$and to term the so-defined
 quantities $\,Q_{\textstyle{_l}}\,$ as the tidal quality factors:
 \ba
 \frac{1}{Q_{\textstyle{_l}}(\chi_{\textstyle{_{lmpq}}})}~=~\sin |\epsilon_{\textstyle{_l}}(\omega_{\textstyle{_{lmpq}}})|~=\sin\epsilon_{\textstyle{_l}}(\chi_{\textstyle{_{lmpq}}})~~.
 \label{B3}
 \ea
 In this notation,
 \ba
 k_{\textstyle{_l}}(\omega_{\textstyle_{lmpq}})~\sin\epsilon_{\textstyle{_l}}(\omega_{\textstyle_{lmpq}})
 \;=\,~ \frac{\textstyle k_{\textstyle{_l}}(\chi_{\textstyle_{lmpq}})}{Q_{\textstyle{_l}}(\chi_{\textstyle_{lmpq}})}~\,\mbox{Sgn}\,\omega_{\textstyle_{lmpq}}
 \quad.
 \label{}
 \ea

 \subsection{Constitutive equation}

 Rheological properties of a material are encoded in a constitutive equation interconnecting the present-time deviatoric strain tensor $\,u_{\gamma\nu}(t)\,$ with the values that have been assumed by the deviatoric stress $\,{\sigma}_{\gamma\nu}(t\,')\,$ over the time $\,t\,'\,\leq\,t\,$. Under linear deformation, the equation has the form of convolution, in the time domain:
 \begin{eqnarray}
 2\,u_{\gamma\nu}(t)\,=\,\hat{J}(t)~\sigma_{\gamma\nu}\,=\,\int^{t}_{-\infty}\stackrel{\;\centerdot}{J}(t-t\,')~
 {\sigma}_{\gamma\nu}(t\,')\,d t\,'~~,~~~
 \label{I12_4}
 \label{E1}
 \end{eqnarray}
 and the form of product, in the frequency domain:
 \begin{eqnarray}
 2\;\bar{u}_{\gamma\nu}(\chi)\,=\;\bar{J}(\chi)\;\bar{\sigma}_{\gamma\nu}(\chi)\;\;.
 \label{LLJJKK}
 \label{E2}
 \end{eqnarray}
 Here $\,\bar{u}_{\gamma\nu}(\chi)\,$ and $\,\bar{\sigma}_{\gamma\nu}(\chi)\,$ are the Fourier images of strain and stress, while the complex compliance $\,\bar{J}(\chi)\,$
 is a Fourier image of the kernel $\,\dot{J}(t-t\,')\,$ of the integral operator (\ref{I12_4}). See, e.g., Efroimsky (2012$\,$a,$\,$b) for details.

 \subsection{Solids. The Maxwell model and its generalisation\label{solids}\label{section3.2}}

 At low frequencies, deformation of most solids is viscoelastic and obeys the Maxwell model. This model can be represented with a viscous damper and an elastic spring connected in series. Experiencing the same force, these elements have their
  elongations summed up. This illustrates a situation where a deviatoric stress $\,{\mathbb{S}}\,$ generates a deviatoric strain comprising a purely elastic part
  $\,\stackrel{(e)}{\mathbb{U}}\,$ and a purely viscous part $\,\stackrel{(v)}{\mathbb{U}}\;$:
 \ba
 {\mathbb{U}}\,=\;\stackrel{(e)}{\mathbb{U}}\,+\,\stackrel{(v)}{\mathbb{U}}
  ~,\,\quad\mbox{where}\,\quad
 \stackrel{(e)}{\mathbb{S}}\,=\,2\,\mu\,\stackrel{(e)}{\mathbb{U}}~~~~\mbox{and}\quad~
 \stackrel{(v)}{\mathbb{S}}\,=\,2\,\eta\,\frac{\partial\,}{\partial t}~\stackrel{(v)}{\mathbb{U}}
  ~~~,
 \label{dddt}
 \ea
 $\,\eta\,$ and $\,\mu\,$ denoting the viscosity and unrelaxed rigidity.

 As in the Maxwell regime both parts of the strain are generated by the same stress
 \ba
 {\mathbb{S}}\;=\;\stackrel{(v)}{\mathbb{S}}\,=\,\stackrel{(e)}{\mathbb{S}}~~~,
 \label{}
 \ea
 the formulae (\ref{dddt}) can be assembled into one equation:
 \begin{subequations}
 \ba
 \stackrel{\centerdot}{\mathbb{U}}\,=\,\frac{1}{2\,\mu}\;\stackrel{\centerdot}{\mathbb{S}}\,+\,\frac{1}{2\,\eta}\;{\mathbb{S}}~~~
 \label{}
 \ea
 or, equivalently:
 \ba
 \stackrel{\centerdot}{\mathbb{S}}\,+\;\frac{1\;}{\tau_{_M}}\,{\mathbb{S}}
 ~=~2\,\mu\,\stackrel{\centerdot}{\mathbb{U}}~~~,
 \label{}
 \ea
 \label{these}
 \end{subequations}
 with the {\emph{Maxwell time}} introduced as
 \ba
 \tau_{_M}\;\equiv\;\frac{\,\eta\,}{\,\mu\,}~~~.
 \label{Maxwell}
 \ea
 Comparing the equation (\ref{these}) with the general expression (\ref{I12_4}) for the compliance operator, we see that the kernel is a time derivative of the compliance function
 \ba
 ^{\textstyle{^{(Maxwell)}}}J(t\,-\,t\,')\,=\,\left[\,J\,+\,\left(t\;-\;t\,'\right)\;\frac{1}{\eta}\,\right]\;\Theta(t\,-\,t\,')~~~,
 \label{Max}
 \ea
 $\Theta(t\,-\,t\,')\,$ being the Heaviside step function, and $\,J\,$ being the unrelaxed compliance:
 \ba
 J~\equiv~\frac{\,1\,}{\,\mu\,}\quad.
 \label{}
 \ea
 In the frequency domain, the equation (\ref{these}) can be cast into the form (\ref{LLJJKK}), with the complex
 compliance given by
 \begin{eqnarray}
 ^{\textstyle{^{(Maxwell)}}}{\bar{\mathit{J\,}}}(\chi)~=~J\,-\,\frac{i}{\eta\chi}~=~J\,\left(\,1~-~\frac{i}{\chi\,\tau_{_M}}\right)\quad,
 \label{don}
 \label{LL42}
 \label{E8}
 \end{eqnarray}
 the terms $\,J\,$ and $\,-\,{i}/(\eta\chi)\,$ giving the elastic and viscous parts of deformation, correspondingly. Such a body becomes elastic at high and viscous at low frequencies.

 It is often convenient to employ the dimensionless versions of the compliance:
 \begin{eqnarray}
 ^{\textstyle{^{(Maxwell)}}}{\bar{\cal{J\,}}}(\chi)\,=
 {^{\textstyle{^{(Maxwell)}}}{\bar{\mathit{J\,}}}(\chi)}\,J^{-1}
 \,=\,1\,-\,i\,(\chi\,\tau_{_M})^{-1}\,~,~~
 \label{E9}
 \label{maxwell}
 \end{eqnarray}
 and of its real and imaginary parts:
 \bs
 \ba
 {^{\textstyle{^{(Maxwell)}}}}\Re(\chi)&=&1~~,
 \label{E39a}\\
 \nonumber\\
  {^{\textstyle{^{(Maxwell)}}}}\Im(\chi)&=&-\,(\chi\,\tau_{_M})^{-1}~~.\qquad~
 \label{E39b}
 \ea
 \label{E39}
 \es
 The Maxwell rheology is a special case of the Andrade rheology that comprises inputs from elasticity, viscosity, and inelastic processes (mainly, dislocation unjamming):
 \begin{subequations}
 \begin{eqnarray}
 {\bar{\mathit{J\,}}}(\chi)&=&J\,+\,\beta\,(i\chi)^{-\alpha}\;\Gamma\,(1+\alpha)\,-\,\frac{i}{\eta\chi}
  \label{112_1}
  \label{E3a}\\
 \nonumber\\
 &=& J\,+\,\beta\,(i\chi)^{-\alpha}\;\Gamma\,(1+\alpha)\,-\,i\,J\,(\chi\,\tau_{_M})^{-1}
 ~~,
 \label{112_2}
 \label{E3b}
 \end{eqnarray}
 where $\,\Gamma\,$ is the Gamma function,  while $\,\alpha\,$ and $\,\beta\,$ denote the dimensionless and dimensional Andrade parameters.

  While conventional, the expressions (\ref{112_1} - \ref{112_2}) for the Andrade compliance suffer an inconvenient feature, the fractional dimensions of the parameter $\,\beta\,$. It was therefore suggested in Efroimsky (2012$ \, $a,$ \, $b) to shape the compliance into a more usable form
 \begin{eqnarray}
 {\bar{\mathit{J\,}}}(\chi)~=~J\,\left[\,1\,+\,(i\,\chi\,\tau_{_A})^{-\alpha}\;\Gamma\,(1+\alpha)~-~i~(\chi\,\tau_{_M})^{-1}\right]\,\;,~~~
 \label{112_3}
 \label{E3c}
 \end{eqnarray}
 \label{LL44}
 \end{subequations}
 with the parameter $\,\tau_{_A}\,$ introduced via
 \begin{eqnarray}
 \beta\,=\,J~\tau_{_A}^{-\alpha}~~.
 \label{beta}
 \label{E4}
 \end{eqnarray}
 and named as $ \, ${\it{the Andrade time}}$ \, $. The compliance (\ref{E3c}) is identical to (\ref{112_1}) and  (\ref{112_2}), but is spared of the parameter of fractional dimensions.

 In the dimensionless form, the Andrade complex compliance reads as:
 \begin{eqnarray}
 {\bar{\cal{J\,}}}(\chi)\equiv\frac{{\bar{\mathit{J\,}}}(\chi)}{J}~=~1\,+\,(i\,\chi\,\tau_{_A})^{-\alpha}\;\Gamma\,(1+\alpha)~-~i~(\chi\,\tau_{_M})^{-1}\quad,
 \label{E5}
 \end{eqnarray}
 its real and complex parts being
 \ba
 \nonumber
 \Re(\chi)&\equiv&{\cal R}{\it e} [ \bar{\cal{J}}(\chi)]~=~\frac{{\cal R}{\it e} [ \bar{J}(\chi)]}{J}~=~
 \quad\qquad\qquad\qquad
 ~\\
 \nonumber\\
 &=&1\;+\;(\chi\tau_{_A})^{-\alpha}\;\cos\left(\,\frac{\alpha\,\pi}{2}\,\right)
 \;\Gamma(\alpha\,+\,1)~~,\qquad
 \label{A3b}
 \label{E6}
 \ea
 \ba
 \nonumber
 \Im(\chi)&\equiv&{\cal I}{\it m} [ \bar{\cal{J}}(\chi)]~=~
 \frac{{\cal I}{\it m} [ \bar{J}(\chi)]}{J}
 \quad\qquad\qquad\qquad\qquad\quad
 ~\\
 \nonumber\\
 &=&-\,(\chi\,\tau_{_M})^{-1}\,-\,(\chi\,\tau_{_A})^{-\alpha}\;\sin\left(
 \,\frac{\alpha\,\pi}{2}\,\right)\;\Gamma(\alpha\,+\,1)~~.\qquad~
 \label{A4b}
 \label{E7}
 \ea

 At frequencies below some threshold (Karato \& Spetzler 1990, Eqn. 17), dislocation unjamming becomes less efficient a process, and the mantle response becomes purely viscoelastic. This means that the Andrade time $\,\tau_{_A}\,$ rapidly increases (so the Andrade term in the formulae (\ref{LL44}) falls off rapidly) when the frequency $\,\chi=\chi_{\textstyle{_{lmpq}}}\,$ goes below the threshold value. As a result of this, $ \, ${\it{at low frequencies}}$ \, $ the complex compliance of solids approaches that of a Maxwell body, and we obtain the formulae (\ref{E8} - \ref{E39}).

 Mind that in this section the rigidity and compliance assume their $\,${\it{unrelaxed}}$\,$ values:
 \ba
 \mu~=~\mu(0)\qquad\mbox{and}\qquad J~=~J(0)\quad.
 \label{}
 \ea

 \subsection{Rubble piles}

 Ample observational data on rotation of asteroids have led to a consensus that many (possibly, most) asteroids are loose aggregates kept together by gravity forces.
 Rheology of such objects should be borrowed from the recent studies on mechanics of granular matter. Such studies render essentially nonlinear constitutive equations which contain parameters other than rigidity or viscosity. We however expect that averaging of such equations over a volume much larger than a typical size of a granule should furnish linearisable equations with $\,${\it{effective}}$\,$ rigidity and viscosity and, possibly, other effective rheological parameters.
 \footnote{~Even if we are aware of the existence and relevance of some rheological parameter, its exact physical meaning is inseparable from context, i.e., from a particular way how the parameter enters the constitutive equation. For example, the same couple of parameters, rigidity and viscosity, enter the Maxwell and Kelvin-Voigt models in different ways. Accordingly, the physical meaning of viscosity within these two models is different. Likewise, different is the meaning of rigidity in these models. Enough to say, the viscous limit is approached through $\,\mu\rightarrow\infty\,$, in the Maxwell model, and through $\,\mu\rightarrow 0\,$, in the Kelvin-Voigt model.\vspace{2mm}
 ~\\
 In their often cited elegant work, Goldreich and Sari (2009) calculated the effective rigidity of a granular aggregate. Although a big step forward, that project still requires continuation. It is necessary to carry out a similar estimate for the effective viscosity of such a medium. An even bigger challenge will be to study how exactly these two (and, possibly, other) parameters enter the constitutive equation.} We leave this part of work for future.

 In this paper, our goal is simply to demonstrate that tidal evolution of an asteroidal binary is parameterised not only by its components' rigidity, but also by other rheological parameters -- like the viscosity.

 Partial justification of treating rubble as a viscoelastic material comes from numerical modeling -- see, e.g., the work by Walsh et al. (2008) who studied both monodisperse (same-size) piles and simple bimodal distributions (those containing two different sizes of particles). It was found that monodisperse aggregates behave in a manner distant from fluid, while mixing of spheres of different sizes yields a closer-to-fluid behaviour. Within a somewhat different model by Tanga et al. (2009), even monodisperse piles
 demonstrated hydrodynamical behaviour. At the same time, simulated rubble piles did not behave like perfect fluids, due to their ability to sustain shear stress.
 (Also see the experimental study by Murdoch et al. 2013.) This means that the aggregates have a finite shear rigidity $\,\mu\,$ and may be regarded viscoelastic.

  \section{Why we need complex Love numbers,\\
  not just $\,k_l\,$ and $\,Q\,$}\label{compla}\label{3.3}

 It has long become common in the astronomical literature to introduce, separately, a degree-$l\,$ Love number $\,k_l\,$ and the corresponding quality factor $\,Q\,$.
 A minor flaw in this tradition is that, fundamentally, the inverse of $\,Q\,$ is the sine of the degree-$l\,$ phase lag: $\,1/Q\,=\,\sin\epsilon_l\,$, so we better equip it with an appropriate subscript: $\,Q_l\,$. A bigger inconsistency is that $\,k_l\,$ and $\,Q_l\,$ cannot be introduced independently, because they both are determined by  self-gravitation and rheology of the body. As we shall see below, this circumstance becomes crucial in description of spin dynamics of asteroids. The only path to consistent treatment lies in introducing complex Love numbers.

 \subsection{Static tide. Love numbers\label{4.1}}

 Let a homogeneous spherical incompressible primary of radius $\,R\,$ be perturbed by a point-like exterior stationary secondary called {\it{perturber}}.
 Introduce a spherical coordinate system associated with the primary. The latitudes $\,\phi,\,\phi^*\,$ are reckoned from the primary's equator, while the longitudes $\,\lambda,\,\lambda^*\,$ are reckoned from a fixed meridian. The perturber of mass $\,M\,$ is stationarily located in an exterior point $\,{\erbold}^{~*}=(r^*,\,\lambda^*,\,\phi^*)\,$.

 Consider an arbitrary exterior point $\,\erbold = (r,\lambda,\phi)\,$ and the surface point $\,\Rbold = (R,\lambda,\phi)\,$ located beneath it (so $\,R\leq r,\,r^*\,$).
 In the surface point $\,\eRbold\,$, the disturbing potential due to the perturber located in $\,\erbold^{\,*}\,$ can be written as
 \ba
 W(\eRbold\,,\,\erbold^{~*})&=&\sum_{{\it{l}}=2}^{\infty}~W_{\it{l}}(\eRbold\,,~\erbold^{~*})\quad,\quad\mbox{with}\quad W_{\it{l}}(\eRbold\,,~\erbold^{~*})\propto P_{\it{l}}(\cos \gamma)\quad,\quad
 \label{W}
 \ea
 $P_{\it{l}}(\cos \gamma)\,$ being the Legendre polynomials, and
 $\,\gamma\,$ being the angle between the vectors ${\erbold}^{\,*}$ and $\Rbold$ originating from the primary's centre.

  Likewise, the additional tidal potential $\,U(\erbold
  ,\,\erbold^{\,*}
 )\,$ in the exterior point $\,\erbold\,$, due to the primary's deformation caused by the perturber being in $\,\erbold^{\,*}\,$,
 can be expanded as:~\footnote{~Since the points $\,\erbold\,$ and $\,\eRbold\,$ are located on the same ray pointing out of the centre of the body, the angle between $\,\erbold^{\,*}\,$ and $\,\erbold\,$ is equal to the angle $\,\gamma\,$ between $\,\erbold^{\,*}\,$ and $\,\eRbold\,$.}
  \ba
 U(\erbold
  \,,\,\erbold^{~*}
 )&=&\sum_{{\it{l}}=2}^{\infty}~U_{\it{l}}(\erbold
  \,,~\erbold^{~*}
 )\quad,\quad\mbox{with}\quad U_{\it{l}}(\erbold
  \,,~\erbold^{~*}
 )\propto P_{\it{l}}(\cos \gamma)\quad.
 \label{U}
 \ea

 A degree-$l\,$ spherical component $\,U_l(\erbold,\,\erbold^{\,*})\,$ in the point $\,\erbold\,$ is proportional to the degree-$l\,$ spherical component $\,W_l(\Rbold\,,\,\erbold^{\;*})\,$ of the perturbing potential in $\,\eRbold\;$:
 \ba
 U_{\it l}(\erbold,\,\erbold^{\,*})=\left(\frac{R}{r}\right)^{l+1}{k}_{l}\;W_{\it{l}}(\eRbold\,,\,\erbold^{\;*})~~.
 \label{}
 \ea
 Then the additional exterior tidal potential of the primary can be written down as:
  \ba
 U(\erbold,\,\erbold^{\,*})~=~\sum_{{\it l}=2}^{\infty}~U_{{l}}(\erbold,\,\erbold^{\,*})~=~\sum_{{\it l}=2}^{\infty}\,\left(\,\frac{R}{r}\,\right)^{{\it l}+1}
 k_{l}\;\,W_{\it{l}}(\eRbold\,,\;\erbold^{\;*})\quad.\quad
 ~~~~~~~~~~~~~~~~
 \label{2}
 \label{L2}
 \ea
 The quantities $\,k_{l}\,$ are the static Love numbers. Solution of a boundary-value problem for potentials, deformation, and stress renders the following expression:
 \ba
 k^{\textstyle{^{(static)}}}_{\it l}\,=\;\frac{3}{2\,({\it l}\,-\,1)}\;\,\frac{1}{1\;+\;{\cal{A}}^{\textstyle{^{(static)}}}_{l}}~=~\frac{3}{2\,({\it l}\,-\,1)}\;\,\frac{1}{1\;+\;{\cal{B}}_{l}\,\mu}~=~\frac{3}{2\,({\it l}\,-\,1)}\;\,\frac{1}{1\;+\;{\cal{B}}_{l}/J}
 ~~~,\qquad
 \label{A_def}
 \ea
 where
 \ba
 {\cal{B}}_{l}\,\equiv~\frac{\textstyle{(2\,{\it{l}}^{\,2}\,+\,4\,{\it{l}}\,+\,3)}}{\textstyle{{\it{l}}\,\mbox{g}\,
 \rho\,R}}~=\;\frac{\textstyle{3\;(2\,{\it{l}}^{\,2}\,+\,4\,{\it{l}}\,+\,3)}}{\textstyle{4\;{\it{l}}\,\pi\,
 G\,\rho^2\,R^2}}~
 \qquad
   ,\quad \label{B} \ea and \ba
 {\cal{A}}^{\textstyle{^{(static)}}}_{\it l}\equiv~{\cal{B}}_{l}\,\mu~=~{{\cal{B}}_{l}}/{J}\quad.\qquad
 \label{stat}
 \ea
 In the above formulae, $\,\rho\,$, $\,$g$\,$, and $\,R\,$ are the density, surface gravity, and radius of the body, while $\,G\,$ signifies the gravitational constant.  Since {\it{static}}$\,$ implies: {\it{relaxed}}, then the rigidity modulus $\,\mu\,$ and the compliance $\,J\equiv1/\mu\,$  assume their $\,${\it{relaxed}}$\,$ values:
 \ba
 \mu~=~\mu(\infty)\qquad\mbox{and}\qquad J~=~J(\infty)\quad,\qquad\mbox{in the static case only}.
 \label{}
 \ea
 This will not be the case for evolving tide to be considered below. There, $\,\mu\,$ and $\,J\,$ will assume their unrelaxed values -- like in Section  \ref{solids}.

  To draw this section to a close, we would mention that in the denominator of (\ref{A_def}) the term $\,1\,$ comes from self-gravitation, while the term $\,{\cal{A}}^{\textstyle{^{(static)}}}_l\,$ originates owing to rheology. This way, $\,{\cal{A}}^{\textstyle{^{(static)}}}_l\,$ serves as a dimensionless measure of strength-dominated versus gravity-dominated static behaviour; see Section \ref{no} below.

  \subsection{Evolving tide. Love operators and complex Love numbers\label{sub}}

 In the case of a perturber ``feeling" the tide it itself creates, we set $\,\erbold=\erbold^{\,*}\,$. This enables us to use a shorter notation $\,U(\erbold)\,$ instead of $\,U(\erbold,\,\erbold^{\,*})\,$. When the tide is evolving, the time-dependence has to be introduced, both in the overall tidal potential $\,U(\erbold,\,t)\,$ and in its Legendre components $\,U_l(\erbold,\,t)\,$.

 Under evolving tidal disturbance, the expression (\ref{L2}) becomes a linear operator (called the {\it{Love operator}}):
 \ba
 U_{l}(\erbold,\,t)\;=\;\left(\frac{R}{r}
 \right)^{l+1}\int_{-\infty}^{t} {\bf\dot{\it{k}}}_{\textstyle{_l}}(t-t\,')~W_{l}
 (\eRbold\,,\;\erbold^{\;*},\;t\,')\,dt\,'~,
 \label{chuk}
 \ea
 where the Love functions $\,k_{\textstyle{_l}}(t-t\,')\,$ are analogues to the compliance functions $\,J(t-t\,')\,$ introduced in the equation (\ref{E1}). Overdot stands for time derivative.

 In the frequency domain, the convolution (\ref{chuk}) becomes
 \footnote{~The equation (\ref{VR}) is simplified, in that a more accurate treatment would render the spectral components as functions of the tidal mode $\,\omega=\omega_{\textstyle{_{lmpq}}}\,$ and not of the physical frequency $\,\chi= \chi_{\textstyle{_{lmpq}}}=\,|\,\omega_{\textstyle{_{lmpq}}}\,|\,$.
 Fortunately, though, the simplification is affordable. As explained in Appendix \ref{AppendixA4} below, the employment of the positive definite physical frequencies instead of the tidal modes is legitimate, insofar as we do not forget to include the sign multiplier ~Sgn$\,\omega_{\textstyle{_{lmpq}}}\,$ into an $\,lmpq\,$ term of the expansion for the torque. With this stipulation in mind, in this article we shall always expand over $\,\chi\,$.}
 \ba
 \bar{U}_{\textstyle{_{l}}}(\chi)\;=\;\left(\,\frac{R}{r}\,\right)^{l+1}\bar{k}_{\textstyle{_{l}}}(\chi)\;\,\bar{W}_{\textstyle{_{l}}}(\chi)\;\;\;,
 \label{VR}
 \label{L33}
 \ea
 with $\,\chi\,$ being the frequency. Here $\,\bar{U}_{\textstyle{_{l}}}(\chi)\,$ and $\,\bar{W}_{\textstyle{_{l}}}(\chi)\,$ are the Fourier or Laplace components of the potentials $\,{U}_{\textstyle{_{l}}}(t)\,$ and $\,{W}_{\textstyle{_{l}}}(t)\,$, while the complex Love numbers $\,\bar{k}_{\textstyle{_{l}}}(\chi)\,$ are the Fourier or Laplace components of the kernels $\,{\bf\dot{\it{k}}}_{\textstyle{_l}}(t-t\,')\,$.

 The frequency-dependencies $\,\bar{k}_{\textstyle{_{l}}}(\chi)\,$ should be derived from the expression for the complex compliance $\,\bar{J}(\chi)\,$ or from its inverse, the complex rigidity $\,\bar{\mu}(\chi)=1/\bar{J}(\chi)\,$. Into that derivation, also the size and mass of the body will enter. So the tidal response will be defined by both rheology and gravity.

 Referring the reader to Efroimsky (2012a,b) and the literature cited therein, we state that for viscoelastic bodies the complex Love number $\,\bar{k}_{l}(\chi)\,$ is related to the complex rigidity $\,\bar{\mu}(\chi)\,$ by the same
 algebraic expression through which the static Love number $\,{k}_{l}\,$ is related to the relaxed rigidity $\,{\mu}(\infty)\,$. Equivalently, one can say that the complex Love number $\,\bar{k}_{l}(\chi)\,$ is related to the complex compliance $\,\bar{J}(\chi)\equiv 1/\bar{\mu}(\chi)\,$ by the same algebraic expression through which the static Love number $\,{k}_{l}\,$ is related to the relaxed compliance $\,{J}(\infty)\equiv 1/{\mu}(\infty)\;$:
 \footnote{~This algebraic similarity is but one manifestation of a more general theorem known as the
 $\,${\it{correspondence principle}}$\,$ or the $\,${\it{elastic-viscoelastic analogy}}$\,$.
 Although the discovery of this principle is sometimes attributed to Biot (1954), an early version thereof can be found in the cornerstone work by Sir George Darwin (1879).}
 \ba
 \bar{k}_{\it l}(\chi)~=~\frac{3}{2\,({\it l}\,-\,1)}\;\,\frac{\textstyle 1}{\textstyle 1\;+\;{\cal{B}}_{l}\;\bar{\mu}(\chi)}
 ~=~\frac{3}{2\,({\it l}\,-\,1)}\;\,\frac{\textstyle 1}{\textstyle 1\;+\;{\cal{B}}_{l}/\bar{J}(\chi)}
 ~~~.~\quad~
  \label{k2bar}
 \ea

 Writing the degree-$l\,$ complex Love number as
 \ba
 \bar{k}_{\it{l}}(\chi)\;=\;{\cal{R}}{\it{e}}\left[\bar{k}_{\it{l}}(\chi)\right]\;+\;\inc\;
 {\cal{I}}{\it{m}}\left[\bar{k}_{\it{l}}(\chi)\right]\;=\;|\bar{k}_{\it{l}}(\chi)|\;
 e^{\textstyle{^{-\inc\epsilon_{\it l}(\chi)}}}
 \label{L36}
 \ea
 we conventionally denote the phase as $\,-\,\epsilon_l\,$, with a ``minus" sign. This convention imparts $\,\epsilon_l\,$ with the meaning of phase lag, as can be understood from the relation (\ref{VR}). Equating the absolute values of the right- and left-hand sides of the relation (\ref{VR}), we make another useful observation: the role of the dynamical Love number is played by the absolute value of the complex Love number:
 \ba
 {k}_{l}~=~|\bar{k}_{l}(\chi)|\quad.
 \label{}
 \ea
 Finally, we notice that
 \ba
 |\bar{k}_{\it{l}}(\chi)|\;\sin\epsilon_{l}(\chi)\;=\;-\;{\cal{I}}{\it{m}}\left[\,\bar{k}_{\it{l}}(\chi)\,
 \right]
 \;\;\;,
 \label{ggffrr}
 \ea
 whence, using the expression (\ref{k2bar}), we arrive at
 \bs
 \ba
 \bar{k}_{\it{l}}(\chi)\;\sin\epsilon_{l}(\chi)~=~-~\frac{3}{2(l-1)}~\frac{{\cal{B}}_{\textstyle{_l}}\;{\cal{I}}{\it{m}}\left[\bar{J}(\chi)\right]
 }{\left({\cal{R}}{\it{e}}\left[\bar{J}(\chi)\right]+{\cal{B}}_{\textstyle{_l}}\right)^2+\left({\cal{I}}{\it{m}}
 \left[\bar{J}(\chi)\right]\right)^2}
 \label{E10a}
 \ea
 where $\,k_l(\chi)\equiv|\bar{k}_{\it{l}}(\chi)|\,$ and, as ever, $\,\chi\,$ is a shortened notation for $\,\chi_{\textstyle{_{lmpq}}}\,$.

 We see that the product $~k_l(\chi_{\textstyle{_{lmpq}}})/Q_l(\chi_{\textstyle{_{lmpq}}})\,\equiv\,k_{\textstyle{_l}}(\chi_{\textstyle{_{lmpq}}})~\sin\epsilon_{\textstyle{_l}}(\chi_{\textstyle{_{lmpq}}})\;$
 standing in an $\,lmpq\,$ term of the expansion for the tidal torque can be calculated consistently through the formulae (\ref{k2bar} - \ref{E10a}), with the coefficients $\,{\cal{B}}_l\,$ rendered by (\ref{B}). For a homogeneous incompressible sphere, the only information needed to calculate this product is: the radius, the density, and the rheological law $\,\bar{J}(\chi)\,$.

 The above expression can be cast into the form of
 \ba
 \bar{k}_{\it{l}}(\chi)\;\sin\epsilon_{l}(\chi)=~-~\frac{3}{2(l-1)}~\;\frac{{\cal{A}}_{\textstyle{_l}}\;\Im(\chi)
 }{\left(\,\Re(\chi)\,+\,{\cal{A}}_{\textstyle{_l}}\,\right)^2\,+\,\Im(\chi)^{\,2}} \quad,\qquad
 \label{E10b}
 \ea
 \label{E10}
 \es
 where
  \ba
 \Re(\chi)\equiv{\cal{R}}{\it{e}}
 \left[\bar{J}(\chi)\right]/J\quad,\qquad\,\Im(\chi)\equiv {\cal{I}}{\it{m}} \left[\bar{J}(\chi)\right]/J\quad,
 \label{}
 \ea
 are the real and imaginary components of the $\,${\it{dimensionless}}$\,$ complex compliance, while
 \ba
 {\cal{A}}_l\,\equiv\,{\cal{B}}_l\,\mu\,=\,{\cal{B}}_l/J
 \label{}
 \ea
 is the dimensionless rigidity. The dimensionless rigidity $\,{\cal{A}}_l\,$ relates to $\,{\cal{B}}_l\,$ in the same way as it used to do in the static case, equation (\ref{stat}). Mind, though, that now $\,\mu\,$ and $\,J\,$ denote the $\,${\it{unrelaxed}}$\,$ values of the rigidity and compliance, not the relaxed ones as in the static problem.
 So the values of $\,{\cal{A}}_l\,$ are different from $\,{\cal{A}}^{\textstyle{^{(static)}}}_l\,$.

 While the product
   $~k_l(\chi_{\textstyle{_{lmpq}}})/Q_l(\chi_{\textstyle{_{lmpq}}})\,\equiv\,k_{\textstyle{_l}}(\chi_{\textstyle{_{lmpq}}})~\sin\epsilon_{\textstyle{_l}}(\chi_{\textstyle{_{lmpq}}})\;$
 shows up in an $\,lmpq\,$ term of the tidal torque, a product
 \ba
 k_{\textstyle{_l}}(\chi_{\textstyle{_{lmpq}}})~\cos\epsilon_{\textstyle{_l}}(\chi_{\textstyle{_{lmpq}}})\;=\;{\cal{R}}{\it{e}}\left[\bar{k}_l(\chi_{\textstyle{_{lmpq}}})\right]
 \label{}
 \ea
 emerges in an $\,lmpq\,$ term of the expansion for the tidal potential. For reference purposes, we provide in Appendix \ref{App} a general expression for this product, as well as its versions for the Maxwell and other rheologies.

 Do draw this section to a close, it would not hurt to mention that the expression (\ref{k2bar}), as well as the ensuing formulae (\ref{E10}), are inapplicable to fluid celestial
 bodies.$\,$\footnote{~Tidal response of fluid objects (like Jupiters or stars) $\,${\it{cannot}}$\,$ be treated as a viscous limit of the Maxwell rheology. Dissipation in these bodies is ruled by the equations of hydrodynamics, with turbulence taken into account (Ogilvie \& Lin 2004; Remus et al. 2012; Auclair-Desrotour et al. 2015).}

 \subsection{The tidal lag versus the ``seismic" lag}

 From the expression (\ref{k2bar}), we obtain the actual dynamical Love numbers:
 \ba
 {k}_{l}(\chi)\,\equiv\;|\,\bar{k}_l(\chi)\,|
 &=&\frac{3}{2\,({\it l}\,-\,1)}\;\sqrt{
 \frac{\Re(\chi)^2+\Im(\chi)^2
 }{\left(\,\Re(\chi)+{\cal{A}}_{\textstyle{_l}}\right)^2+\Im(\chi)^2}
 }
 ~~~.~\quad~
 \label{epsilon}
 \ea
 The ratio of the formulae (\ref{E10b}) and (\ref{epsilon}) gives us the inverse tidal quality factors:
 \ba
 Q_l(\chi)^{-1}\;\equiv\;\sin\epsilon_l(\chi)\;=\;-\;\frac{
 {\cal{A}}_{\textstyle{_l}}\,\Im(\chi)
 }{
 \sqrt{\Re(\chi)^2+\Im(\chi)^2\,}~
 \sqrt{\left(\,\Re(\chi)+{\cal{A}}_{\textstyle{_l}}\right)^2+\Im(\chi)^2}\,
 }
 \label{}
 \ea

 The tidal quality factors should not be confused with the ``seismic" quality factor whose inverse is the sine of the phase lag $\,\delta\,$ showing up in the rheological law:
 \ba
 Q(\chi)^{-1}\;\equiv\;\sin\delta(\chi)~=~-\;\frac{\Im(\chi)}{\sqrt{\Re(\chi)^2+\Im(\chi)^2\,}}\quad.
 \label{delta}
 \ea
 The former and latter formulae interrelate the tidal and ``seismic" phase lags:
 \ba
 \sin\epsilon_l(\chi)\;=\;\frac{{\cal{A}}_{\textstyle{_l}}}{\sqrt{\left(\,\Re(\chi)+{\cal{A}}_{\textstyle{_l}}\right)^2+\Im(\chi)^2}\,}~\sin\delta(\chi)\quad.
 \label{comparison}
 \ea
 To the best of our knowledge, this relation has never appeared in the literature hitherto.

\begin{figure}[htbp]
 \vspace{2.5mm}
 \centering
 \includegraphics[angle=0,width=0.68\textwidth]{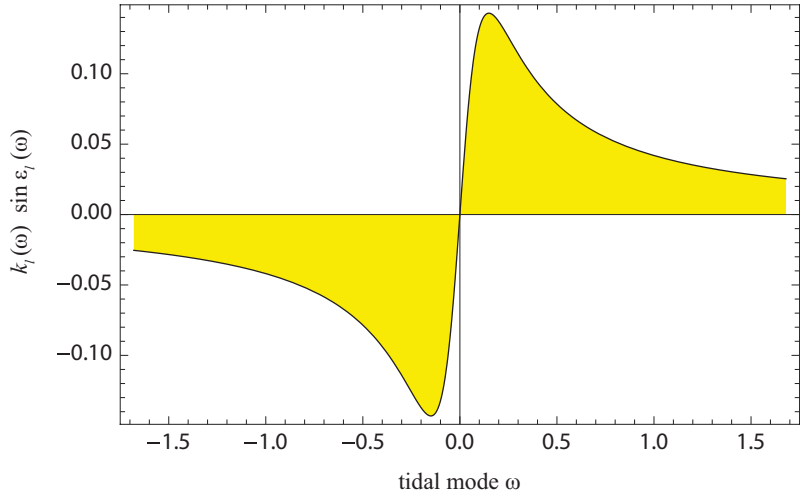}
 \caption{\small{~A typical shape of the quality function $\,k_l(\omega)\,\sin\epsilon_l(\omega)\,$, ~where $\,\omega\,$ is a shortened notation for the tidal Fourier mode
 $\,\omega_{\textstyle{_{lmpq}}}\,$. ~~(From Noyelles et al. 2014.)
 \label{Fig1}}}
 \end{figure}

 As one would expect, the lags coincide when rheology ``beats" gravity -- i.e., when $\,{\cal{A}}_{\textstyle{_l}}\,\gg\,|\,\Re(\chi)\,|\,,~\,|\,\Im(\chi)\,|\,$.
 Otherwise, the tidal phase lags deviate considerably from the nominal ``seismic" lag $\,\delta\,$. This happens on approach to a spin-orbit resonance, i.e., to a zero frequency. While the ``seismic" lag stays constant (and approaches $\,\pi/2\,$, in the case of Maxwell body), the tidal lags continuously go through zero -- and so do the products $\,k_{\textstyle{_l}}\,\sin\epsilon_{\textstyle{_l}}\,$, as can be seen from Figure \ref{Fig1}. In this figure, we depicted $\,k_{\textstyle{_l}}\,\sin\epsilon_{\textstyle{_l}}\,$ not as a function of the positive definite physical frequency $\,\chi=\chi_{\textstyle{_{lmpq}}}\,$, but as a function of the Fourier mode $\,\omega=\omega_{\textstyle{_{lmpq}}}\,$ that can have either sign.$\,$\footnote{~Recall that $\,\chi_{\textstyle{_{lmpq}}}=\,|\,\omega_{\textstyle{_{lmpq}}}\,|\,$.}
 The shape of this function is about the same for all linear rheologies. For a Maxwell body, the extrema are located at
 \ba
 \omega_{peak}\,=\;\pm\;\frac{1}{\tau_{_M}\,{\cal{B}}_{l}\;\mu}\;=\;\pm\;\frac{\textstyle{8\,\pi\,
 G\,\rho^2\,R^2}}{\textstyle{57\;\eta}}\quad.
 \label{peak}
 \ea
 The peaks' amplitude is virtually insensitive to our choice of the viscosity $\,\eta\,$. At the same time, the spread between the extrema depends on $\,\eta\,$. As we see from the formula (\ref{peak}), for a higher viscosity the peaks are located close to the point of resonance. As the viscosity assumes smaller values, the peak frequency spreads out, eventually reaching and bypassing the orbital frequency $\,n\,$. In realistic settings, however, this would require extremely low viscosities.

 Outside the inter-peak interval, the function $\,k_l(\omega)\,\sin\epsilon_l(\omega)\,$ behaves as the inverse $\,Q\,$ in seismology, and changes very slowly with frequency.

 Between the peaks, the products $\,k_l(\omega)\,\sin\epsilon_l(\omega)\,$ are almost linear in
 $\,\omega=\omega_{\textstyle{_{lmpq}}}\,$, and go continuously through zero. This enables the tidal
 torque to transcend spin-orbit resonances continuously (Makarov \& Efroimsky 2013,
 Makarov et al. 2012, Noyelles et al. 2014).

 It can be demonstrated that the linearity of $\,k_l\,\sin\epsilon_l\,$ in $\,\omega\,$ is equivalent to the time lag being frequency independent: $\,\Delta t_l(\omega_{\textstyle{_{lmpq}}})\,=\,\Delta t\,$, see Efroimsky \& Makarov
 (2013).$\,$\footnote{~Some authors call the constant-$\Delta t$ tidal model $\,${\it{viscous}}$\,$. This is not a good term, since in the viscous limit of the Maxwell model we still arrive at a kink shape of $\,k_l(\omega)\,\sin\epsilon_l(\omega)\,$, $\,$like in Figure \ref{Fig1}. The CTL (constant
 time lag) tidal model is a better term for this approach.} This means that the tidal response of an actual terrestrial body can be approximated with the constant-$\Delta t\,$  model $\,${\it{only when all considered tidal frequencies are lower than}} $\,|\,\omega_{peak}\,|\;$
 -- $\,$or, roughly speaking, when all mean motions and spin rates are lower than $\,|\,\omega_{peak}\,|\,$.
 To achieve this, we would have to endow the viscosity with very low values. This observation absolutely prohibits the application of the CTL (constant time lag) tidal model to solid or semi-molten silicate planets (while for liquified planets this entire formalism is not intended anyway).

 \section{Complex Love numbers for different rheologies}

 For homogeneous near-spherical objects, the formulae (\ref{E10}) provide a turn-of-the-crank method. Inserting therein the rheology $\,\bar{J}(\chi)\,$, along with the values of the radius and density, we obtain the functions $~k_l(\chi)/Q_l(\chi)\,\equiv\,k_{\textstyle{_l}}(\chi)~\sin\epsilon_{\textstyle{_l}}(\chi)\;$, with $\,\chi=\chi_{\textstyle{_{lmpq}}}\,$. The complex compliance $\,\bar{J}(\chi)\,$ is a function of the tidal frequency and rheological parameters. This is why the tidal torque is sensitive both to the frequency and the rheological parameters (and, thereby, also to the temperature). A similar conclusion is valid for the tidal heating rate (Henning et al. 2009, Makarov \& Efroimsky 2014).

 \subsection{Self-gravitating Maxwell sphere}

 Insertion of the complex compliance (\ref{don}) into the formulae (\ref{k2bar}) and (\ref{E10}) furnishes:
 \bs
 \ba
 \bar{k}_{\textstyle{_l}}(\chi)&=&\frac{3}{2(l-1)}\;\frac{1}{\;\textstyle 1\;+~\frac{\textstyle
 \tau_{\textstyle{_M}}\;\inc\;\chi}{\textstyle \;1\;+\;\tau_{\textstyle{_M}}\;\inc\;\chi\;}\;\,{\cal{B}}_{\textstyle{_l}}\,\mu\;}
 \label{150a}\\
 \nonumber\\
 &=&\frac{3}{2(l-1)}\;\frac{1}{\;\textstyle 1\;+\;\frac{\textstyle
 \tau_{\textstyle{_M}}\;\inc\;\chi}{\textstyle 1\;+\;\tau_{\textstyle{_M}}\;\inc\;\chi\;}~\,{\cal{A}}_{\textstyle{_l}}\;\;}
 \label{150b}
 \ea
 \label{150}
 \es
 and
 \bs
 \ba
 \frac{k_l(\chi)}{Q_l(\chi)}\,\equiv\,k_{\textstyle{_l}}(\chi)~\sin\epsilon_{\textstyle{_l}}(\chi)&=&\frac{3}{2(l-1)}\;\frac{{\cal{B}}_l~\frac{\textstyle 1}{\textstyle \eta\,\chi}}{~(J\,+\,{\cal{B}}_l)^2\,+~\frac{\textstyle 1}{\textstyle \eta^2\,\chi^2}~}
 \label{151a}\\
 \nonumber\\
 &=&\frac{3}{2(l-1)}\;\frac{{\cal{A}}_l~\frac{\textstyle 1}{\textstyle \tau_{_M}\,\chi}}{~(1\,+\,{\cal{A}}_l)^2\,+~\frac{\textstyle 1}{\textstyle \tau_{_M}^2\,\chi^2}~}\quad.\qquad~
 \label{151b}
 \ea
 \label{151}
 \es
 %
 Given by (\ref{B}), the factors $\,{\cal{B}}_l\,$ are dimensional, while  $\,{\cal{A}}_l=\,{\cal{B}}_l\,\mu\,=\,{\cal{B}}_l/J\,$ are dimensionless. Recall that, in distinction from the static case, here $\,\mu\,$ and $\,J\equiv1/\mu\,$ denote the $\,${\it{unrelaxed}}$\,$ (not static) values of the rigidity and compliance, wherefore our $\,{\cal{A}}_{l}\,$ should not be misidentified with their counterparts $\,{\cal{A}}^{\textstyle{^{(static)}}}_{l}$ from Section \ref{4.1}.

 Generalisation of the formula (\ref{151}) to the Andrade model can be found in Efroimsky (2012$\,$a, $\,$eqn. 170).

 \subsection{Three extreme cases\label{ext}}

 \subsubsection{Self-gravitating drop of an ideal fluid. ~No rheology at all\label{no}}

 In the ``no-rheology" limit, the material becomes an ideal fluid wherein shear deformation entails no stress. This implies:
 \begin{subequations}
 \ba
 &~&
 \eta\rightarrow 0\quad,\qquad
 \mu\rightarrow 0\quad,\qquad J\equiv \frac{\,1\,}{\,\mu\,}\rightarrow \,\infty\quad,   \qquad\qquad
 \label{}\\
 \nonumber\\
 &~&\mbox{wherefrom}\qquad\bar{J}(\chi)\rightarrow \infty\qquad\mbox{and}\qquad{\cal{A}}_{\textstyle{_l}}\rightarrow 0\quad,\qquad\qquad\qquad
 \label{}
 \ea
 \end{subequations}
 though the real part of the $\,${\it{dimensionless}}$\,$ compliance stays finite: $\,\Re(\chi)~\rightarrow~1\,$.
 Thence,
 \ba
 \bar{k}_l(\chi)~\rightarrow~\frac{3}{2(l-1)}\quad,
   \label{a} \ea \ba
 k_{\textstyle{_l}}(\chi
 )~\sin\epsilon_{\textstyle{_l}}(\chi
 )~\rightarrow~0
 \quad.
 \label{E2}
 \ea
 From (\ref{a}) we see that in this limit the complex Love number becomes real. Due to the absence of rheology as such, this value of the Love number originates solely from self-gravitation. From (\ref{E2}) we understand that all terms in the expression for torque become zero. No heat is produced either.

 \subsubsection{Self-gravitating elastic body}

 This limit is implemented through
 \begin{subequations}
 \ba
 \tau_{_M}\rightarrow \infty\quad,\qquad\eta\rightarrow\infty\quad,\qquad\mbox{with}~\,J\,~\mbox{staying}~\,\mbox{finite}~~.\quad
 \label{}
 \ea
 Thence the complex compliance becomes real:
 \ba
  \bar{J}(\chi) ~ \rightarrow ~ J\quad.
 \label{}
 \ea
 \end{subequations}
 This again renders the complex Love numbers real:
 \ba
 \bar{k}_{\textstyle{_l}}(\chi
 )\;\rightarrow\;
 \frac{\textstyle 3}{\textstyle 2\,(l-1)}~\frac{\textstyle 1}{\textstyle 1\,+\,{\cal{B}}_{\textstyle_{l}}\,\mu}
 ~=~
 \frac{\textstyle 3}{\textstyle 2\,(l-1)}~\frac{\textstyle 1}{\textstyle 1\,+\,{\cal{A}}_{\textstyle_{l}}}\quad,
 \label{200}
 \ea
 \ba
 k_{\textstyle{_l}}(\chi
 )~\sin\epsilon_{\textstyle{_l}}(\chi
 )~\rightarrow~0
 \quad.\qquad\qquad\qquad\qquad\qquad\qquad\quad
 \label{18}
 \ea
 Phase lags vanish, no tidal torque emerges, and no dissipation is taking place.

 \subsubsection{Self-gravitating perfectly viscous body\label{visc}}

 In this situation, we have to get rid of rigidity by setting $\,\mu\rightarrow\infty\,$ ~(yes, $\,\infty\,$)$\,$:
 \ba
 &~&
 \tau_{_M}\rightarrow 0\quad,\qquad
 \mu\rightarrow \infty\quad,\qquad J\equiv \frac{\,1\,}{\,\mu\,}\rightarrow \,0\quad,   \qquad\qquad
 \label{asbest}\\
 \nonumber\\
 &~&\mbox{wherefrom}\qquad\bar{J}(\chi)\rightarrow ~-~\frac{i}{\eta\chi}\qquad\mbox{and}\qquad{\cal{A}}_{\textstyle{_l}}\rightarrow\infty\quad.\qquad\qquad\qquad
 \label{}
 \ea
 In this limit, the expressions (\ref{150a}) and (\ref{151a}) become:
 \ba
 \bar{k}_{\textstyle{_l}}(\chi)~\rightarrow ~\frac{3}{2(l-1)}\;\frac{1}{\;\textstyle 1\;+~\textstyle
 i~\eta~\chi\;\,{\cal{B}}_{\textstyle{_l}}\;}\quad, ~\qquad~\qquad\qquad~\qquad~
 \label{160}
 \ea
 \ba
 k_{\textstyle{_l}}(\chi)~\sin\epsilon_{\textstyle{_l}}(\chi)~
 \rightarrow ~\frac{3}{2(l-1)}~\frac{{\cal{B}}_{\textstyle{_l}}\;\frac{\textstyle 1}{\textstyle \eta\,\chi}
 }{{\cal{B}}_{\textstyle{_l}}^{\,2}~+~\frac{\textstyle 1}{\textstyle \eta^2\,\chi^2}}
 ~=~~\frac{3}{2(l-1)}~\frac{{\cal{B}}_{\textstyle{_l}}\;{\textstyle \eta\,\chi}
 }{{\cal{B}}_{\textstyle{_l}}^{\,2}\,\eta^2\,\chi^2~+~1}
  \quad.
 \label{161}
 \ea
 It should however be noted that the viscous limit in these formulae cannot be interpreted too literally. Recall that all our algebra rests on the cornerstone fact that the complex $\,\bar{k}_{\textstyle{_l}}(\chi)\,$ relates to the complex $\,\bar{J}(\chi)\,$ in the same manner as their static counterparts do. This is an approximation that stays valid insofar as the accelerations and inertial forces can be neglected in the equation of motion (see Efroimsky 2012$\,$a$\,$ and references therein). In the absence of rigidity, the Coriolis force comes into the picture and entails complicated hydrodynamics (Ogilvie \& Lin 2004, Auclair-Desrotour et al. 2015). So the compliance $\,J\,$ cannot be set exactly zero. It should sooner be understood that the formulae (\ref{160}) and (\ref{161}) pertain to a finite $\,J\ll\eta\chi\,$. If such highly viscous objects exist in the world of small bodies, then the tidal torque (and dissipation) for these objects is proportional not to the rigidity $\,\mu\,$ but to $\,\eta\chi\,$, as can be seen from the above expression.

 \section{Numbers}

 Given our sparse knowledge of the actual parameters and structure of asteroids and comets, we are not sure how good approximation the Maxwell model is. We shall employ it for the time being, due to the lack of anything better.

  As for the parameters' values, we only know that the mean density of most asteroids is about $\,\rho=2\times 10^{3}\,$ kg/m$^3\,$, up to a factor of 1.5 (Pravec \& Harris 2007), while the density of comets is around $\,0.6\times 10^{3}\,$ kg/m$^3\,$ (Britt et al. 2006).

 \subsection{Rigidity beats gravity -- and goes out}

 Assuming the nominal values $\,\mu_{\textstyle{_{(rock)}}}(\infty)=5\times 10^{10}\,$ Pa and $\,\rho=3.5\times 10^{3}\,$ kg/m$^3\,$, Goldreich \& Sari (2009, Section 3.1) obtain, for rubble piles:
 \bs
 \ba
 {\cal{A}}_2^{\textstyle{^{(static)}}}\propto~10^4\quad\mbox{for}\quad R~\approx~10^4~\mbox{m}\quad,
 \label{}
 \ea
  \ba
 {\cal{A}}_2^{\textstyle{^{(static)}}}\propto~10^2\quad\mbox{for}\quad R~\approx~10^6~\mbox{m}\quad.
 \label{}
 \ea
 \label{}
 \es
 Although the unrelaxed rigidity differs from its relaxed (static) counterpart, for regular solids the difference is not dramatic. We presume that for rubble piles, too, the said difference does not exceed two orders of magnitude. If this presumption is right, we can accept, for the unrelaxed dimensionless rigidity, that
 \ba
 {\cal{A}}_l\,\gg\,1\quad,\quad\mbox{for}\quad R\,\lesssim\,10^6~\mbox{m}~~.
 \label{owing}
 \ea
 This should not come as a surprise, because $\,{\cal{A}}_l\,$ being larger than unity implies that in small bodies ``rigidity beats gravity".

  Now we can safely assume in the formulae (\ref{150} - \ref{151}) that, for all asteroids and comets (and for not too large satellites), $\,1+{\cal{A}}_l\approx {\cal{A}}_l\,$ and, equivalently, $\,J+{\cal{B}}_l\approx {\cal{B}}_l\,$. Hence the following approximation for the squared Love number:
 \ba
 k_{\textstyle{_l}}^2(\chi)~=~\left[\,\frac{3}{2(l-1)}\,\right]^2\frac{1~+~\tau_{_M}^2\,\chi^2}{1~+~\tau_{_M}^2\,\chi^2~(1\,+\,{\cal{A}}_l)^2}~
 \approx~\left[\,\frac{3}{2(l-1)}\,\right]^2\frac{1~+~\eta^2\,\chi^2\,J^2}{1~+~\eta^2\,\chi^2~{\cal{B}}_l^2}\quad,\quad
 \label{170a}
 \ea
 and for $\,k_l/Q_l\;$:
 \ba
 k_{\textstyle{_l}}(\chi)~\sin\epsilon_{\textstyle{_l}}(\chi)~=~\frac{3}{2(l-1)}\;\frac{{\cal{B}}_l~\frac{\textstyle 1}{\textstyle \eta\,\chi}}{~(J\,+\,{\cal{B}}_l)^2\,+~\frac{\textstyle 1}{\textstyle \eta^2\,\chi^2}~}~\approx~\frac{3}{2(l-1)}\;\frac{{\cal{B}}_l~\frac{\textstyle 1}{\textstyle \eta\,\chi}}{~{\cal{B}}_l^2\,+~\frac{\textstyle 1}{\textstyle \eta^2\,\chi^2}~}\quad.\qquad
 \label{170b}
 \ea
 We see that the Love number becomes rigidity-independent for $\,\eta\,\chi\,\ll\,\mu\,$. Moreover, our second entity, $\,k_l/Q_l\,$, does not depend upon rigidity at all.$\,$\footnote{~We can use the equation (\ref{170b}) to express $\,k_l/Q_l\,$ as a function of the dimensionless rigidity $\,{\cal{A}}_l\;$:
 \ba
  k_{\textstyle{_l}}(\chi)~\sin\epsilon_{\textstyle{_l}}(\chi)~=~\frac{3}{2(l-1)}\;\frac{{\cal{A}}_l~\,\frac{\textstyle 1}{\textstyle \tau_{_M}\,\chi}}{~{\cal{A}}_l^2\,+~\frac{\textstyle 1}{\textstyle \tau_{_M}^2\,\chi^2}~}\quad.\qquad~
 \nonumber
 \ea
 However, since $\,{\cal{A}}_l={\cal{B}}_l\,\mu\,$ and $\,\tau_{_M}=\eta/\mu\,$, casting our expression into the above form will only camouflage the absence of an actual dependence upon $\,\mu\,$ or $\,J\,$.}

  The latter fact warrants a seemingly paradoxical conclusion that the rigidity of small bodies is $\,${\it{too large to be relevant}}$\,$ for description of tidal lagging in these bodies; too large -- even with a reduction derived by Goldreich \& Sari (2009). In reality, this conclusion works also for monoliths, because in the expression for $\,{\cal{A}}_l\,$ a higher rigidity of a solid rock is partially compensated by its higher density squared. So the conclusion works for all small bodies, including satellites, -- provided they are (a) not too large, so the inequalities (\ref{owing}) hold; and (b) may be regarded as Maxwell bodies.

 \subsection{Viscosity versus self-gravitation. A practical criterion}

 Our next goal is to examine the denominators of the right-hand sides of the expressions (\ref{170a}) and (\ref{170b}). In both denominators, it is important whether
 $\,{\cal{B}}_l\,\chi\,\eta\,$ is much larger or much smaller than unity. With the expression (\ref{B}) for $\,{\cal{B}}_l\,$ kept in mind, this furnishes us with the following criterion.

 For tidal frequencies and effective viscosities obeying
 \ba
  \frac{57}{8\,\pi}~\frac{\chi~\eta}{G\,\rho^2\,R^2}~\gg~1\quad,
 \label{high}
 \ea
 the expressions for $\,k_l\,$ and $\,k_l/Q_l\,$ read as: \footnote{~Evidently, for even higher viscosities, the Love numbers become simply $\,k_l\approx\,\frac{\textstyle 3 }{\textstyle 2(l-1)}~\frac{\textstyle 1 }{\textstyle {\cal{A}}_{\textstyle{_{l}}}}~$.}
 \ba
 k_{\textstyle{_l}}(\chi)~\approx~\frac{3}{2(l-1)}\;\frac{~\sqrt{1~+~\eta^2\,\chi^2\,J^2}~}{~{\cal{B}}_l~\eta\;\chi}\quad,\quad
 \label{rembrandt}
 \ea
 \ba
 k_{\textstyle{_l}}(\chi)~\sin\epsilon_{\textstyle{_l}}(\chi)~\approx~\frac{3}{2(l-1)}~\frac{\textstyle 1}{\textstyle {\cal{B}}_l\,\eta\,\chi}\quad.\qquad
 \label{rubens}
 \ea
 They are good everywhere except in the low-frequency limit. There, the inequality (\ref{high}) is no longer satisfied, and we have to switch from (\ref{rubens}) to the full expression (\ref{170b}).

  If the product $\,\chi\,\eta\,$ assumes low values,
  we arrive at a situation with  \footnote{~The numerical factor on the left-hand side of (\ref{ll}) corresponds to $\,l=2\,$. If higher degrees are included, the factor should be chosen as $~3\,(2l^2+4l+3)/(4l\pi)~$ for the maximal $\,l\,$ taken into account.}
 \ba
 \frac{57}{8\,\pi}~\frac{\chi~\eta}{G\,\rho^2\,R^2}~\ll~1\quad,
 \label{ll}
 \label{low}
 \ea
 so the expressions of our concern become:
 \ba
 k_{\textstyle{_l}}(\chi)~\approx~\frac{3}{2(l-1)}\quad,\qquad
 \label{degas}
 \ea
 \ba
 k_{\textstyle{_l}}(\chi)~\sin\epsilon_{\textstyle{_l}}(\chi)~\approx~\frac{3}{2(l-1)}\;\,{\cal{B}}_l\;\eta\;\chi\quad.\qquad
 \label{velazquez}
 \ea
 The latter expression can be used to explore tidal entrapment into spin-orbit states.

 \section{Conclusions}

 In the article, we have shown that in asteroidal binaries the intensity of tidal interactions (i.e., the tidal torques and the ensuing spin and orbit evolution) depends primarily on the product of the tidal frequency $\,\chi\,$ by the viscosity $\,\eta\,$. Contrary to a long-entrenched opinion, the rigidity $\,\mu\,$ exerts virtually no influence on tidal evolution.

 We have developed a quantitative criterion that enabled us to separate two types of the functional dependence of $\,k_l/Q_l\,$ upon the said product $\,\chi\eta\,$. For higher values of this parameter, we obtained: $\,k_l/Q\propto 1/(\eta\chi)\;$,  see the equations (\ref{high} - \ref{rubens}).   For lower values of $\,\chi\eta\,$, we got:  $\,k_l/Q\propto \eta\chi\,$, see the formulae (\ref{low} - \ref{velazquez}). The dependencies (\ref{rubens}) and (\ref{velazquez}) determine the spin histories, including tidal captures into spin-orbit resonances.$\,$\footnote{~Be mindful that the approximation (\ref{rubens}) cannot be used in the vicinity of a resonance where $\,\chi\,$ vanishes. In that vicinity, one has to employ the full expression (\ref{170b}) or its approximation (\ref{velazquez}).}

 Our study rests on an assumption that small bodies can be described with the Maxwell model. The model is well applicable to rigid rocks at low frequencies, so our conclusions are safely valid for monoliths. We apply this approach also to rubble, for the lack of a better model so far. As we learn more about mechanics of granular mixtures in a weak gravity field, we should be prepared to amend the tidal theory with other rheological parameters, ones that do not show up in viscoelastic mechanics.

 It should be finally added that the old misbelief about the allegedly key role of the rigidity has pervaded also the literature on inelastic relaxation of precessing
 (wobbling) asteroids and comets (Prendergast 1958; Burns \& Safronov 1973; Efroimsky 2001, 2002; Molina et al. 2003; Ryabova 2004; Sharma et al. 2005; Breiter et al. 2012; Breiter \& Murawiecka 2015).
  In an upcoming work (Frouard \& Efroimsky 2015), we shall demonstrate that in the said setting, too, the leading role belongs to viscosity.

\section*{Acknowledgments}

 I wish to thank Konstantin Batygin, Julien Frouard, Val{\'{e}}ry Lainey and Valeri Makarov for our useful discussions on topics related to this project.


 \appendix
 \section{
 \Large{Appendix\label{appendix}}.\vspace{2mm}
 \label{AppendixA}}

 In this Appendix, we briefly summarise the basic facts from the theory of bodily tides. For more detailed explanation, see Efroimsky \& Makarov (2013) and references therein.

 \subsection{Fourier modes and physical frequencies}

 As we mentioned in Section \ref{section2.1}, both the potential from the perturber and the additional tidal potential of the perturbed body can be expanded over the Fourier modes
 \ba
 \omega_{\textstyle{_{lmpq}}}
  ~\approx~(l-2p+q)\;n\,-\,m\;\dot{\theta}\quad,\qquad
 \label{}
 \ea
 with $\,n\,\equiv\;\stackrel{\bf\centerdot}{\cal{M\,}}$ being the ``anomalistic" mean motion and $\,{\bf{\dot{\theta\,}}}\,$ being the spin rate of the tidally perturbed body. These modes can assume either sign, while the physical frequencies of stresses and strains in the perturbed body are positive definite:
 \ba
 \chi_{\textstyle{_{lmpq}}}\,=\,|\,\omega_{\textstyle{_{lmpq}}}\,|~\approx~|\,(l-2p+q)\;n\,-\,m\;\dot{\theta}\,|\,~.
 \label{}
 \ea

  \subsection{Phase lags and dynamical Love numbers}

 The tidal reaction always lags in time, compared to action. Each Fourier mode $\,\omega_{\textstyle{_{lmpq}}}\,$ has a time lag of its own, $\,\Delta t_l (\omega_{\textstyle{_{lmpq}}})\,$. Owing to causality, time lags are positive definite:
 \ba
 \Delta t_l (\omega_{\textstyle{_{lmpq}}})\;>\;0\quad.
 \label{A4}
 \ea
 For physical reasons, the time lags are defined by the forcing frequencies $\,\chi_{\textstyle{_{lmpq}}}\equiv\,|\,\omega_{\textstyle{_{lmpq}}}\,|\,$, and are ignorant of the sign of the modes $\,\omega_{\textstyle{_{lmpq}}}\;$:
 \ba
 \Delta t_l (\,-\,\omega_{\textstyle{_{lmpq}}})\;=\;\Delta t_l (\omega_{\textstyle{_{lmpq}}})\;=\;\Delta t_l (\chi_{\textstyle{_{lmpq}}})\quad.
 \label{A5}
 \ea
 In the Darwin-Kaula theory of tides,
 the phase lags are products of the modes $\,\omega_{lmpq}\,$ by the corresponding time lags:
 \begin{subequations}
 \ba
 \epsilon_l(\omega_{\textstyle{_{lmpq}}})&=&\omega_{\textstyle{_{lmpq}}}~\,\Delta t_l(\omega_{\textstyle{_{lmpq}}})
 \label{lags_1}
 \label{A6a}\\
 &=&\chi_{\textstyle{_{lmpq}}}~\,\Delta t_l(\chi_{\textstyle{_{lmpq}}})~\,\mbox{Sgn}\,(\,\omega_{\textstyle{_{lmpq}}}\,)\,~.
 \label{lags_2}
 \label{A6b}
 \ea
 \label{lags}
 \label{A6}
 \end{subequations}
 The functional forms of the lags and Love numbers (as functions of the Fourier mode $\,\omega_{lmpq}\,$) are defined only by the degree $\,l\,$. Hence the
 notation: $\,k_{\textstyle{_{l}}}(\omega_{\textstyle{_{lmpq}}})\,$, $\,\epsilon_{\textstyle{_{l}}}(\omega_{\textstyle{_{lmpq}}})\,$, $\,\Delta t_{\textstyle{_{l}}}(\omega_{\textstyle{_{lmpq}}})\,$. This, however, is
 true only for a homogeneous spherical body.~\footnote{~For oblate homogeneous bodies, the functional dependencies of the lags and Love numbers on modes are defined by both the degree $\,l\,$ and order $\,m\,$ (Dehant 1987, Ogilvie 2013). So one should write: $\,k_{\textstyle{_{lm}}}(\omega_{\textstyle{_{lmpq}}})\,$, $\,\epsilon_{\textstyle{_{lm}}}(\omega_{\textstyle{_{lmpq}}})\,$, and $\,\Delta t_{\textstyle{_{lm}}}(\omega_{\textstyle{_{lmpq}}})\,$. These Love numbers and lags differ from the Love numbers and lags of the spherical reference body by terms of the order of the flattening, so a small non-sphericity can be neglected.\label{F}
 }

  Keeping in mind that the time lag $\,\Delta t_{\textstyle{_l}}(\omega_{\textstyle{_{lmpq}}})\,$ is positive definite, we see from (\ref{lags_1}) that the signs of  $\,\epsilon_l\,$ and $\,\omega_{\textstyle{_{lmpq}}}\,$ always coincide. So the phase lags are odd functions of $\,\omega_{lmpq}\,$ and thus can be written as
 \begin{subequations}
 \begin{eqnarray}
 \epsilon_{\textstyle{_l}}(\omega_{\textstyle{_{lmpq}}})~=~|\,\epsilon_{\textstyle{_l}}(\omega_{\textstyle{_{lmpq}}})\,|\,~\mbox{Sgn}\,(\omega_{\textstyle{_{lmpq}}})
  \label{epsa}
  \label{A7a}
  \ea
  or, equivalently, as
  \ba
  \epsilon_{\textstyle{_l}}(\omega_{\textstyle{_{lmpq}}})~=~\epsilon_{\textstyle{_l}}(\chi_{\textstyle{_{lmpq}}})~\,\mbox{Sgn}\,(\omega_{\textstyle{_{lmpq}}})~~,
 \label{epsb}
 \label{A7b}
 \end{eqnarray}
 \label{eps}
 \label{A7}
 \end{subequations}
 with $~\epsilon_{\textstyle{_l}}(\chi_{\textstyle{_{lmpq}}})\,=\,|\,\epsilon_{\textstyle{_l}}(\omega_{\textstyle{_{lmpq}}})\,|~$ being positive-definite.

 Just as the time lags, the dynamical Love numbers depend on the actual physical frequencies $\,\chi_{\textstyle{_{lmpq}}}\,$. This is the same as to say that they are even functions of $\,\omega_{\textstyle{_{lmpq}}}\;$:
 \begin{eqnarray}
 k_{\textstyle{_l}}(\,-\,\omega_{\textstyle{_{lmpq}}})~=~k_{\textstyle{_l}}(\omega_{\textstyle{_{lmpq}}})~=~k_{\textstyle{_l}}(\chi_{\textstyle{_{lmpq}}})~~.
 \label{k}
 \label{A8}
 \end{eqnarray}
 From the latter two equations, it ensues that
 \begin{eqnarray}
 k_{\textstyle{_l}}(\omega_{\textstyle{_{lmpq}}})~\sin\epsilon_{\textstyle{_l}}(\omega_{\textstyle{_{lmpq}}})~=~
    k_{\textstyle{_l}}(\chi_{\textstyle{_{lmpq}}})~\sin\epsilon_{\textstyle{_l}}(\chi_{\textstyle{_{lmpq}}})~\mbox{Sgn}\,(\omega_{\textstyle{_{lmpq}}})\,~.
 \label{ksin}
 \label{A9}
 \end{eqnarray}

 \subsection{Observable geometric lags}

 Geometrically, tides comprise a system of superimposed bulges moving around the body in different directions, at different angular, and with different lags. The principal, semidiurnal bulge has an elliptic shape. As seen from the surface, it moves at the same angular rate as the perturber, with a positive or negative lag.
 Lesser bulges have more involved shapes and different velocities. By geodetic measurements, it is possible to single out those bulges, and to measure their magnitudes and geometric lags (Benjamin et al. 2006, Eanes \& Bettadpur 1996).

 The geometric lag angle of an $\,lmpq\,$ bulge is (Efroimsky \& Makarov 2013, Eqn 26):
 \ba
 \delta_{lmpq}\,=~\frac{\omega_{lmpq}}{m}~\Delta t_l(\omega_{lmpq})\,~.
 \label{geometric}
 \ea
 For example, the geometric lag of the principal bulge is $\,\delta_{2200}=\,\frac{\textstyle \omega_{2200}}{\textstyle 2}~\Delta t_2\,=\,(n-\dot{\theta})\,\Delta t_2\,$, where the time lag is taken at the appropriate, semidiurnal mode: $\,\Delta t_2=\Delta t_2(\omega_{2200})\,$.
 Also mind that the tide components with $\,m=0\,$ are longitude-independent and the notion of geometric lag becomes meaningless -- see the equation (25) in {\it{Ibid}}.

 \subsection{Why positive frequencies are insufficient\label{AppendixA4}}

 While the general formula for a Fourier expansion of a field includes integration or summation over both positive and negative modes, in the case of real fields it is sufficient to expand over positive frequencies only.~\footnote{~The condition of a field being real requires that the real part of a Fourier term at a negative mode is equal to the real part of the term at an opposite, positive mode. Thus one can get rid of the terms with negative modes, at the cost of doubling the appropriate terms with positive modes. (The convention is that the actual field is the real part of a complex expression.)\vspace{1.5mm}} Surprisingly, the tidal theory makes a rare exception from this rule. This theory cannot be expressed via the physical frequencies $\,\chi_{\textstyle{_{lmpq}}}\,=\,|\,\omega_{\textstyle{_{lmpq}}}\,|\,$ solely, with no mentioning of the modes $\,\omega_{\textstyle{_{lmpq}}}\,$. It turns out that a contribution from a Fourier mode into the additional tidal potential is $\,${\it{not}}$\,$ physically equivalent to the contribution from the mode of an opposite sign.

 To understand the reason for this, recall that the expansion for the tidal potential is needed to obtain expressions for tidal force and torque. These are infinite series too; and in each of these two series an $\,lmpq\,$ term contains explicitly the sign of the appropriate tidal mode $\,\omega_{\textstyle{_{lmpq}}}\,$. Specifically, an $\,lmpq\,$ term in the expressions for the tidal force or torque contains a factor
 \begin{eqnarray}
 k_{\textstyle{_l}}(\omega_{\textstyle{_{lmpq}}})~\sin\epsilon_{\textstyle{_l}}(\omega_{\textstyle{_{lmpq}}})~=~
     k_{\textstyle{_l}}(\chi_{\textstyle{_{lmpq}}})~\sin\epsilon_{\textstyle{_l}}(\chi_{\textstyle{_{lmpq}}})~\mbox{Sgn}\,(\omega_{\textstyle{_{lmpq}}})\,~,
  \label{} \end{eqnarray}
 where we made use of the formulae (\ref{epsb}) and (\ref{k}).

 This way, if we choose to expand the tidal torque or the tidal force over the positive definite frequencies $\,\chi_{\textstyle{_{lmpq}}}\,$ only, we shall have to insert ``by hand" the multipliers ~Sgn$\,\omega_{\textstyle{_{lmpq}}}\,$ into each $\,lmpq\,$ term of these expansions (Efroimsky 2012a,b).

 \subsection{A typical shape of the function $~k_{\textstyle{_l}}(\omega_{\textstyle{_{lmpq}}})~\sin\epsilon_{\textstyle{_l}}(\omega_{\textstyle{_{lmpq}}})~$}

 The function $~k_{\textstyle{_l}}(\omega_{\textstyle{_{lmpq}}})~\sin\epsilon_{\textstyle{_l}}(\omega_{\textstyle{_{lmpq}}})\,=\,
 k_{\textstyle{_l}}(\chi_{\textstyle{_{lmpq}}})/Q_{\textstyle{_l}}(\chi_{\textstyle{_{lmpq}}})~\mbox{Sgn}\,(\omega_{\textstyle{_{lmpq}}})~$ enters the $\,lmpq\,$ term in the expression (\ref{4}) for the tidal torque.
 The kink shape of this function is amazingly robust: it is about the same for all linear rheological models, including the most extreme ones (say, purely viscous mantle with no rigidity).
 It will exhibit a sharp kink, as in Figure \ref{Fig1}, with two peaks having opposite signs.
 The fact that the function  goes continuously through zero enables the torque to transcend spin-orbit resonances continuously.

 Close to $\,\omega_{\textstyle{_{lmpq}}}=\,0\,$, the frequency dependence makes two extrema, and demonstrates a slow frequency-dependence at higher frequencies.

 \subsection{Expression for the function $~k_{\textstyle{_l}}(\omega_{\textstyle{_{lmpq}}})~\cos\epsilon_{\textstyle{_l}}(\omega_{\textstyle{_{lmpq}}})~$\label{App}}

 As was mentioned in Section \ref{sub} above, a product
 \ba
 k_{\textstyle{_l}}(\chi_{\textstyle{_{lmpq}}})~\cos\epsilon_{\textstyle{_l}}(\chi_{\textstyle{_{lmpq}}})\;=\;{\cal{R}}{\it{e}}\left[\bar{k}_l(\chi_{\textstyle{_{lmpq}}})\right]
 \label{}
 \ea
 stands in an $\,lmpq\,$ term of the expansion for the tidal potential $\,U\,$.  It is equal to
 \bs
 \ba
 k_{\textstyle{_l}}(\chi_{\textstyle{_{lmpq}}})~\cos\epsilon_{\textstyle{_l}}(\chi_{\textstyle{_{lmpq}}})
 &=&\frac{3}{2(l-1)}~\;\frac{
 \left(\,{\cal{R}}{\it{e}}\left[\bar{J}(\chi)\right]\,+\,{\cal{B}}_{\textstyle{_l}}\,\right)\;{\cal{R}}{\it{e}}\left[\bar{J}(\chi)\right]\,+\,
 \left({\cal{I}}{\it{m}}\left[\bar{J}(\chi)\right]\,\right)^2
 }{
 \left(\,{\cal{R}}{\it{e}}\left[\bar{J}(\chi)\right]\,+\,{\cal{B}}_{\textstyle{_l}}\,\right)^2\,+\,\left({\cal{I}}{\it{m}}\left[\bar{J}(\chi)\right]\,\right)^2
 }\qquad\qquad
 \label{}\\
 \nonumber\\
 \nonumber\\
 &=&\frac{3}{2(l-1)}~\;\frac{
 \left(\,\Re(\chi)\,+\,{\cal{A}}_{\textstyle{_l}}\,\right)\;\Re(\chi)\,+\,\Im(\chi)^{\,2}
 }{
 \left(\,\Re(\chi)\,+\,{\cal{A}}_{\textstyle{_l}}\,\right)^2\,+\,\Im(\chi)^{\,2}
 }
 \quad.
 \label{}
 \ea
 \es
 For a self-gravitating Maxwell sphere, this expression becomes:
 \ba
 k_{\textstyle{_l}}(\chi_{\textstyle{_{lmpq}}})~\cos\epsilon_{\textstyle{_l}}(\chi_{\textstyle{_{lmpq}}})~=~\frac{3}{2(l-1)}~\left(\,1\;-\;\frac{ {\cal{A}}_l }{ {\cal{A}}_l^2\,+\,\frac{\textstyle 1}{\textstyle \chi^2\,\tau^2_{_M}}}\,\right)
 \label{}
 \ea
 Of a special interest are the three extreme cases considered in Section \ref{ext}:
 \ba
 \mbox{self-gravitating ideal-fluid drop}:\qquad k_{\textstyle{_l}}(\chi_{\textstyle{_{lmpq}}})~\cos\epsilon_{\textstyle{_l}}(\chi_{\textstyle{_{lmpq}}})&\rightarrow &
 \frac{3}{2(l-1)}\quad,\qquad\qquad
 \label{}\\
 \mbox{self-gravitating elastic sphere}:\qquad k_{\textstyle{_l}}(\chi_{\textstyle{_{lmpq}}})~\cos\epsilon_{\textstyle{_l}}(\chi_{\textstyle{_{lmpq}}})&\rightarrow &
 \frac{3}{2(l-1)}~\left(\,1\;-\;{\cal{A}}_l^{-1} \,\right)\quad,\qquad\qquad
 \label{}\\
 \mbox{self-gravitating viscous sphere}:\qquad k_{\textstyle{_l}}(\chi_{\textstyle{_{lmpq}}})~\cos\epsilon_{\textstyle{_l}}(\chi_{\textstyle{_{lmpq}}})&\rightarrow &
 \frac{3}{2(l-1)}~\frac{1}{1\,+\,{\cal{B}}_l^{\,2}\,\chi^2\,\tau^2_{_M}}\quad.\qquad\qquad
 \label{}
 \ea
  Recall, however, that the Kaula theory of bodily tides becomes inapplicable in the limit of a perfectly viscous body, as was explained in Section \ref{visc}.

 \label{lastpage}

\end{document}